\documentclass[traditabstract]{aa}
\usepackage{graphicx}
\usepackage{txfonts}
\usepackage{longtable}
\usepackage{url}
\usepackage{wasysym}
\usepackage{natbib}
\bibpunct{(}{)}{;}{a}{}{,}

\begin{document}
\title{\object{HD~144432}: A young triple system\thanks{Based on observations made with ESO Telescopes at the La Silla Paranal Observatory
under programme ID 075.C-0668 and 083.A-9013.}}  %both (A)
\author{A.~M\"uller\inst{\ref{inst1}}
  \and A.~Carmona\inst{\ref{inst2},\ref{inst3}}
  \and M.~E.~van~den~Ancker\inst{\ref{inst4}}
  \and R.~van~Boekel\inst{\ref{inst1}}
  \and Th.~Henning\inst{\ref{inst1}}  
  \and R.~Launhardt\inst{\ref{inst1}}}
\institute{Max Planck Institute for Astronomy, K\"onigstuhl 17, D-69117 Heidelberg, Germany\\\email{amueller@mpia.de}\label{inst1}
  \and ISDC Data Centre for Astrophysics \& Geneva Observatory, University of Geneva, chemin d\textquoteright Ecogia 16, 1290 Versoix, Switzerland\label{inst2}
  \and Observatoire de Gen\`{e}ve,  University of Geneva, chemin des Maillettes 51, 1290 Sauverny, Switzerland\\\email{andres.carmona@unige.ch}\label{inst3}
  \and European Southern Observatory, Karl-Schwarzschild-Str. 2, D-85748 Garching b. M\"unchen, Germany\label{inst4}
  }
\date{Received date / Accepted date}
\abstract{
We present new imaging and spectroscopic data of the young Herbig star HD~144432~A, which is a well-known binary star with a separation of 1.47\arcsec. High-resolution NIR imaging data obtained with NACO at the VLT reveal that HD~144432~B itself is a close binary pair with a separation of 0.1\arcsec. High-resolution optical spectra, acquired with FEROS at the 2.2m MPG/ESO telescope in La Silla, of the primary star and its co-moving companions are used to determine their main stellar parameters, such as effective temperature, surface gravity, radial velocity, and projected rotational velocity by fitting synthetic spectra to the observed stellar spectra. The two companions, HD~144432~B and HD~144432~C, are identified as low-mass T~Tauri stars of spectral type K7V and M1V, respectively. From the position in the HRD, the triple system appears to be co-eval with a system age of 6$\pm$3~Myr.
}
 % - Stars: mass-loss
\maketitle
\section{Introduction}\label{sec:intro}
Herbig Ae/Be \citep[HAeBe,][]{her60,fin84} stars are intermediate-mass (2-10~M$_{\odot}$) pre-main sequence stars (PMS). They are the higher-mass counterparts of the T Tauri stars (TTS), hence fill the parameter space between TTSs and high-mass young stars in addressing the question of star formation as a function of mass. Double or multiple stellar systems are of great interest to star formation theories as they provide constraints on the components of star formation models, such as fragmentation, accretion, N-body dynamics, and orbit migration \citep[e.g.][and references therein]{tok08}.
Although several imaging surveys have indicated that HAeBe stars have close companions \citep[e.g.][]{lei97,pir97,kou05}, relatively few studies have constrained the physical properties of these companions. One interesting question that can be addressed is the age of the HAeBe star by constraining the age of its companions. An independent and more robust age estimate can be derived from the lower-mass companions.

\citet{cor98} pointed out that the companions of Be stars tend to be intermediate-mass stars with spectral types A to F, while the companions of Ae stars are low-mass TTSs with spectral types K to M, which suggests that the mass ratio seems to be quantitatively independent of the primary mass \citep[e.g.][]{hog92}.

HD~144432~A is a late A to early F-type star \citep{the94} at a distance of 160$^{+36}_{-25}$~pc \citep{lee07} that is associated with the Sco~OB2-2 star forming region. The star is surrounded by a protoplanetary disk \citep[e.g.][]{mee01} and displays signs of active mass accretion. The values found in the literature range from 1.8 to 8.5$\cdot10^{-8}$~$\mbox{M$_{\odot}$\,yr$^{-1}$}$ \citep{blo06,gar06,don11}. From J=3-2~\element[][12]{CO} measurements and line profile fitting, \citet{den05} derived a dust mass of 2$\cdot$10$^{-3}$~M$_{\odot}$ (assuming a constant gas-dust ratio of 100) and an outer radius of 60$\pm$20~AU ($i=48^{\circ}\pm10^{\circ}$) for the disk around HD~144432~A where emitting gas is present.
The system HD~144432 is known to be a binary \citep{per04,car07}. \emph{K}-band observations by \citet{per04} revealed a 2.4~mag fainter stellar companion at a separation of 1.4\arcsec~\citep[$PA$=4$\degr$,][]{car07}. \citet{car07} reported the detection of \ion{Li}{I} (6708~\AA) and \element[][]{H}$\alpha$ emission in the spectrum of HD~144432~B using the VLT-FORS2 spectrograph.\\
\citet{per04} showed that HD~144432~A and B share the same common proper motion and age. In addition, component B was ruled out as a background star by \citet{car07}.
\\
\\
In this paper, we present NACO observations of HD~144432, which resolve HD~144432~B as a close binary star, thereby revealing HD~144432 as a hierarchical triple system. Using high-resolution optical spectroscopy, we determined the stellar parameters of all components of the triple system.
\section{Observations and data reduction}\label{sec:obs}
\subsection{NACO}\label{sec:naco}
\begin{figure}
  \centering
  \includegraphics[scale=0.5]{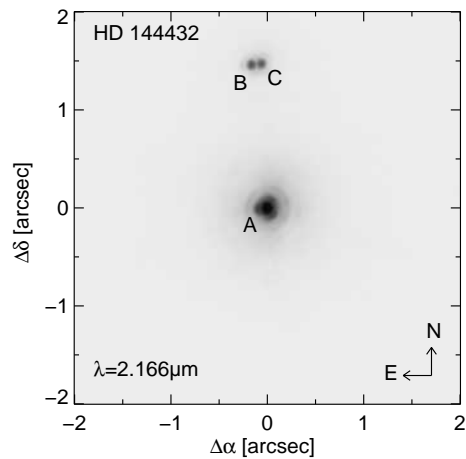}
  \caption{Near infrared image (logarithmic scaling in brightness) of the HD~144432 triple system observed with NACO in July 2005. The components are marked by letters in hierarchical order. The separation between A and B is 1.47\arcsec. The components B and C are separated by 0.1\arcsec. \label{fig:NACO_2166}}
\end{figure}

HD~144432 was observed in the night of July 13, 2005, with NACO \citep{len03,rou03}, the adaptive optics (AO), near-infrared (NIR) camera at ESO's VLT. We obtained these unpublished observations from the ESO archive. The data consist of six imaging observations in three different filters ($K_{s}$ at $\lambda=2.18~\mu$m, NB2.12 at $\lambda=2.122~\mu$m, and NB2.17 at $\lambda=2.166~\mu$m). Each observation consists of eight single exposures at different sky positions (jitter). Standard data reduction (bias subtraction, flat-field correction, bad-pixel interpolation) was applied to each frame by employing the available calibration frames. The average $FWHM$ of the images is 0.07\arcsec. A combined image was produced by applying the shift-and-add method (a two-dimensional cross-correlation routine) to the NB2.17 observation (Fig.~\ref{fig:NACO_2166}) using NACO pipeline recipes provided by ESO\footnote[1]{\url{http://www.eso.org/sci/software/pipelines/naco/naco-pipe-recipes.html}}.
\subsection{FEROS}\label{sec:feros}
We obtained single-epoch spectra of HD~144432~A (SNR of 350 at 5500~\AA) and a combined spectrum of HD~144432~B and C (SNR of 250 at 5500~\AA) in the night of June 1, 2009, using FEROS \citep{kau99} at the 2.2m MPG/ESO telescope at La Silla Observatory in Chile. FEROS covers the optical spectral range from 3600~\AA~to 9200~\AA~and provides a spectral resolution of $\approx$48\,000. The spectra were obtained using the object-calibration mode where one of the two fibers is positioned on the target star and the other fiber is fed with the light of a \element[][]{Th}\element[][]{Ar}+\element[][]{Ne} calibration lamp. 
The reduction of the raw data was performed using the online data reduction pipeline available at the telescope. The pipeline performs bias subtraction and flat-fielding, traces and extracts the single echelle orders, applies the wavelength calibration, and corrects for the barycentric motion. For each exposure, it produces 39 individual sub-spectra representing the individual echelle orders, as well as one merged spectrum.\\
The fiber aperture of FEROS is 2\arcsec~on the sky. The projected separation between the components B and C is 0.1\arcsec. Therefore, the observed spectrum is a superposition of those of the two companions. In addition, owing to the difference in the brightnesses of B and C in the optical with respect to A ($\delta\sim5$ mag) and a seeing at the time of the observations of 0.8\arcsec, the spectrum of the fainter companions is contaminated by the $\approx$1.5\arcsec~apart and $\approx$20 times brighter primary star. 
\section{Data analysis and results}\label{sec:analysis}
\subsection{Imaging}\label{sec:imaging}
To measure the relative position of the companions in the field, i.e. their projected distance and position angle with respect to the primary star, and its relative fluxes, we used the IDL-based program Starfinder \citep{dio00}, which is designed to analyze AO images using the extracted point spread function from the image. The final positional values were derived by averaging the single values of all individual NACO exposures (48 in total). Their errors were derived by computing the standard deviations in the means and taken into account the calibrations of \citet{cha10} for the plate scale (13.25$\pm$0.06~mas) and true north orientation ($-0.02\degr\pm0.10\degr$), which were derived from a data set taken in August 2005. Table~\ref{tbl:astro} lists the measured distances and position angles of the companions.
\begin{table}
  \caption{Relative astrometric measurements of the triple system HD~144432, where $\rho$ is the angular separation, $d$ is the projected separation, and $PA$ is the position angle. \label{tbl:astro}}
  \centering
  \begin{tabular}{lccc}
  \hline\hline
  Parameter & A-B & A-C & B-C \\
  \hline
   $\rho$ / [\arcsec]   & 1.465$\pm$0.007 & 1.468$\pm$0.007 & 0.102 $\pm$ 0.001 \\%angular separation
   $d$ / [AU] & 234$\pm$44      & 235$\pm$44      & 16 $\pm$ 3 \\%projected separation
  $PA$ / [\degr]      & 6.31$\pm$ 0.12  & 2.32$\pm$ 0.12  & 275.90 $\pm$ 0.28 \\
  \hline
  \end{tabular}
\end{table}
The $K_{s}$ magnitudes for B and C were obtained from the relative flux measurements. We computed the zero point of each individual image using HD~144432~A as reference \citep[$K=5.888$~mag,][]{cut03} and derived 9.09~mag for B and 9.16~mag for C, respectively.
\subsection{Spectral classification of HD~144432 A}\label{sec:stellparamsA}
\begin{figure}
  \centering
  \includegraphics[scale=0.4]{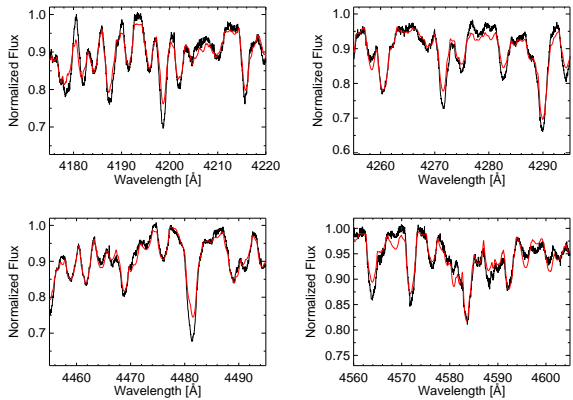}
  \caption{The spectral windows used for fitting the synthetic spectrum (red line) to the observed stellar spectrum (black line) of HD~144432~A.\label{fig:fitexp}}
\end{figure}

\begin{figure}
  \centering
  \includegraphics[scale=0.35]{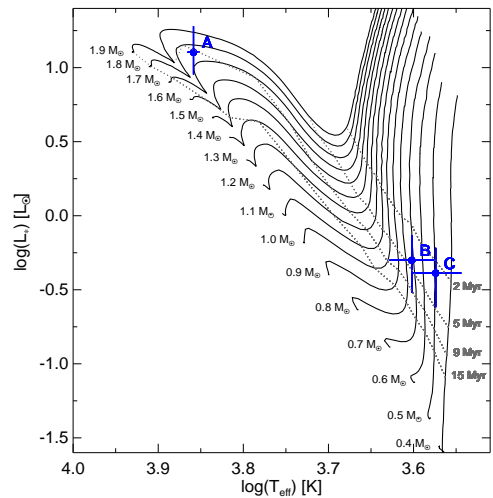}
  \caption{The position of the A, B, and C components of HD~144432 in an HRD along with the evolutionary tracks of \citet{sie00} for a metal abundance of Z=0.02.\label{fig:hrd}}
\end{figure}

Table~\ref{tbl:stellparam} lists the derived stellar parameters of HD~144432~A. The luminosity, $L_{\star}$, was computed by integrating the total flux under the SED. To construct the SED, we used photometric data from \citet{geu90}, \citet{syl96}, and \citet{mal98}, and data from the ISO and IRAS point-source catalog. We derived an extinction $A_{V} = 0.15$ for HD 144432 A by fitting a reddened stellar atmosphere model with the $T_{\rm eff}$ and $\log g$ derived from the spectral fit to the observed UV-optical photometric data. The parameters $T_\mathrm{eff}$, $\log g$, and $v\sin i$ were computed using a self-developed tool for retrieving stellar parameters of Herbig stars based on fitting synthetic spectra to the observed stellar spectrum. The computation of the synthetic spectra was carried out using {\sc spectrum} \citep{gra94} with the {\sc atlas9} atmosphere models \citep{cas04}. The grid of synthetic spectra has step sizes of 125~K in $T_{\rm eff}$, 0.15 in $\log g$, and 5~km\,s$^{-1}$ in $v\sin i$. The stellar spectrum was divided into four spectral windows (Fig.~\ref{fig:fitexp}). Each window was fitted independently and the individual best fits were averaged to derive the final stellar parameters (Table~\ref{tbl:stellparam}). For a detailed description of this procedure, we refer to a forthcoming paper (M\"{u}ller et al., in preparation). 
After finding the best-fit value for $T_\mathrm{eff}$ and $\log g$ at solar metallicity, we tried models with diverse metallicities to find the best fit to the \ion{Fe}{I} lines. We found that the optimal match to the spectrum was provided by solar metallicity. Figure~\ref{fig:fitexp} shows four related windows of the observed stellar spectrum of HD~144432~A (black line), which is closely fitted by a single star synthetic spectrum (red line).\\
We measured the stellar radial velocity, $RV$, by cross-correlating the stellar spectrum with a template spectrum, which was a synthetic spectrum representing the stellar parameters of HD~144432~A (Table~\ref{tbl:stellparam}). 
The resulting cross-correlation function was fitted by a Gaussian function. The position of the center of the Gaussian yields $RV$.
\\
From the position of HD~144432~A in the HRD (Fig.~\ref{fig:hrd}), we obtained the stellar mass, radius, and its age using the evolutionary tracks of \citet{sie00}. We also compared with the evolutionary tracks of \citet{tog11}, which yield the same results. The derived parameters are in close agreement with values derived in other publications, e.g., \citet{dun97}, \citet{boe05}, and \citet{gui06}. The measured effective temperature and surface gravity are comparable to those of an F0IIIe main-sequence star.
\begin{table}
  \caption{Stellar parameters of HD~144432~A, B, and C. \label{tbl:stellparam}}
  \centering
  \begin{tabular}{lccc}
  \hline\hline
  Component & A & B & C\\
\hline
  Parameter & \multicolumn{3}{c}{Value} \\
\hline
  Spectral type & F0IIIe & K7V & M1V \\
  $L_{\star}$ / [L$_{\odot}$] & 12.7 $^{+6.4}$ $_{-3.7}$ & 0.50 $^{+0.24}$ $_{-0.20}$ & 0.41 $^{+0.19}$ $_{-0.16}$ \\
  $T_\mathrm{eff}$ / [K] & 7220$\pm$115 & 4000$\pm$250 & 3750$\pm$250 \\
  $\log g$ / [cm\,s$^{-2}$] & 3.60$\pm$0.15 & 4.0$\pm$0.5 & 4.0$\pm$0.5 \\
  $M_{\star}$ / [M$_{\odot}$] & 1.8 $^{+0.2}$ $_{-0.1}$ & 0.8$\pm$0.2 & 0.5$\pm$0.2 \\
  $R_{\star}$ / [R$_{\odot}$] & 2.3$\pm$0.5 & 1.5$\pm$0.7 & 1.5$\pm$0.7 \\
  $v\sin i$ / [km\,s$^{-1}$] & 75.0$\pm$3.5 & 55$\pm$5 & 40$\pm$5 \\
  $RV$ / [km\,s$^{-1}$] & -2.8$\pm$0.4 & -5$\pm$2 & -4$\pm$2 \\
  $age$ / [Myr] & 9$\pm$2 & 4 $^{+5}$ $_{-2}$ & 3 $^{+2}$ $_{-1}$ \\ %9 $^{+1}$ $_{-2}$
\hline
  \end{tabular}
\end{table}
\subsection{Spectral classification of HD~144432 B and C}\label{sec:stellparamsBC}
The observed spectrum of HD~144432 B and C is strongly contaminated by HD~144432~A. The contamination varies as a function of wavelength, with the blue part of the spectrum being the most strongly affected. In addition, the spectra of components B and C are superimposed. To derive the spectral type of B and C, we compared the continuum-normalized B+C observed spectrum with a continuum-normalized synthetic spectrum made of three components: a high-resolution Kurucz theoretical spectra of B and C (computed as described in Sec.~\ref{sec:stellparamsA}), and a scaled observed FEROS spectrum of HD~144432~A.

We employed a custom designed interactive IDL software\footnote[2]{IDL widget software available upon request from A. Carmona} that permitted us to visually compare in real time, after any change of parameters, the observed and the synthetic spectra. The procedure consisted first choosing the spectral types of B and C thereby automatically fixing their absolute magnitudes ($M_V$); the $M_V$ for the contribution of component A was then varied to match strength of the features of A; and finally, the $v\sin i$ for components B and C was set to reproduce the widths of the absorption lines. The synthetic comparison spectrum was constructed by scaling the flux of each component by its $M_V$, summing the scaled fluxes, and normalizing the resultant spectrum by dividing it by a second order polynomial fit to the continuum. The $\chi^{2}_{\rm red}$ statistic was calculated for each synthetic spectrum.

We used as a first guess a K4V spectrum for B and C \citep{car07}, and searched for a region in the spectrum best suited to the spectral classification. We selected the region at 7695 -- 8100~\AA~ because: ({\it i}) it has few atmospheric lines; ({\it ii}) it is relatively featureless in the spectrum of A; ({\it iii}) it has a very strong line at 7700~\AA~that allows us to differentiate spectral types K and M; ({\it iv}) it has several additional weaker absorption lines that aid the spectral classification and determination of $RV$ and $v\sin i$; and ({\it v}) it has a strong absorption line at 7775~\AA~from component A that is not blended with absorption lines of K and M stars.

After defining the spectral window for classification, we searched for the optimal spectral type pair able to reproduce most of the spectral features observed. The best match was given by a binary system composed of a K7V ($T_{\rm eff}$=4000~K, $\log g$=4.0) and an M1V ($T_{\rm eff}$=3750~K, $\log g$=4.0) star of $v\sin i$=55$\pm$5 and 40$\pm$5~km\,s$^{-1}$, respectively, plus a contribution by A of similar magnitude as the K7V component. We checked other regions of the spectrum (6020--6420~\AA, 6330--6740~\AA, 6655--6760~\AA, 7250--7585~\AA) in addition to the 7695 -- 8100~\AA~region, and confirmed that this binary pair provided a good match (we note that the level of A is different for each region). The uncertainty in the classification is one spectral sub-class and $\Delta\log g$=0.5, owing to the 250~K and 0.5 grid, used for $T_{\rm eff}$ and $\log g$.

\begin{figure}
  \centering
  \includegraphics[scale=0.25]{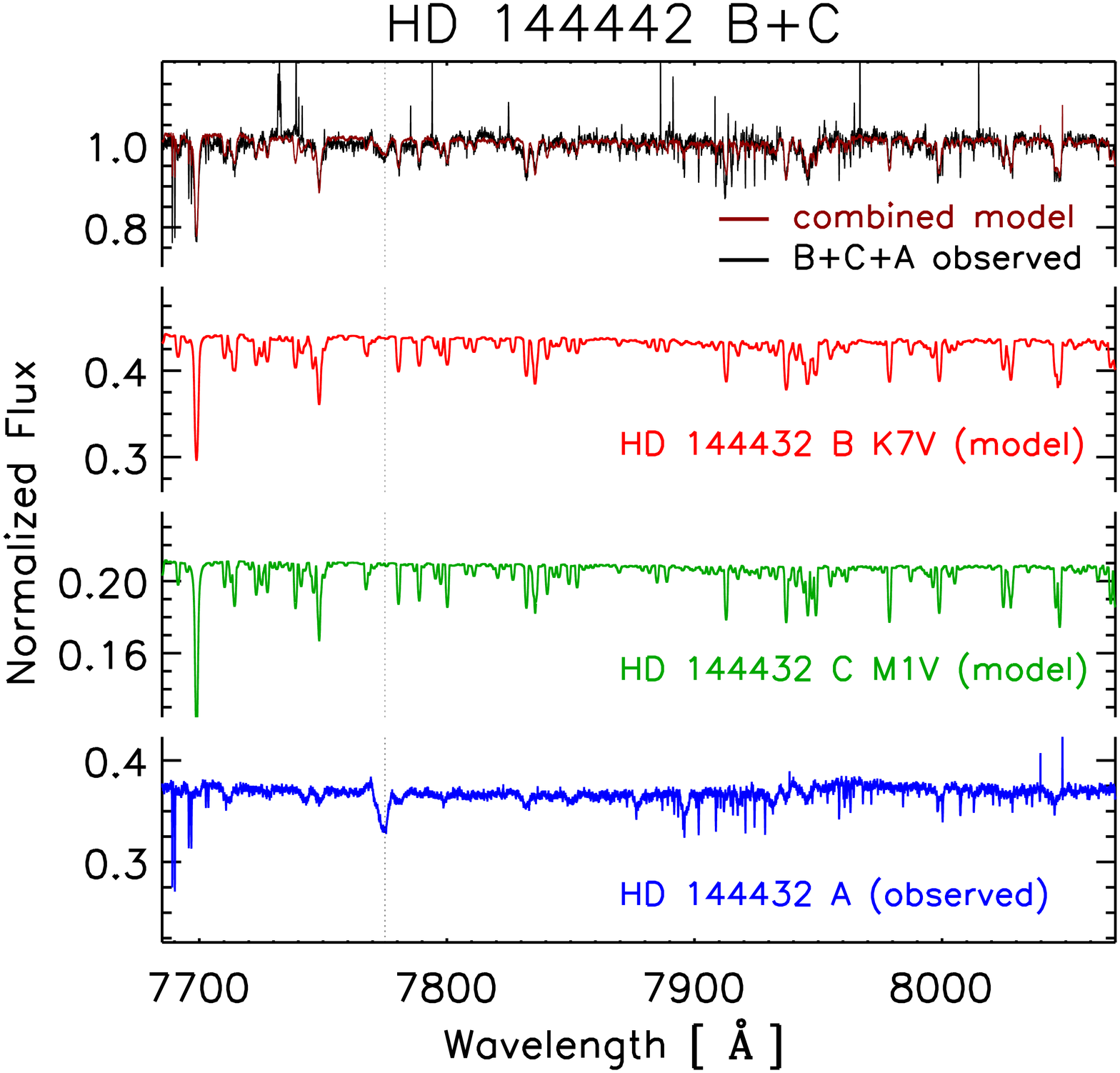}
  \caption{{\it Upper spectra:} observed normalized FEROS spectrum at the position  HD~144432 B and C (in black), and the normalized synthetic combined model that most closely describes the observed spectrum (in red).
  {\it Lower spectra:}
  Relative contributions of HD~144432 B, C, and the contamination from HD~144432 A (note the difference in the Y-axis for each spectrum). 
  The vertical dotted line displays the absorption line of HD~144432 A used to scale the flux of HD~144432 A to match the observed spectrum. 
  The strength of the contamination from the A component is similar to the flux of HD~144432 B.\label{fig:HD144432BC}}
\end{figure}

Once the spectral types and $v\sin i$ of B and C were derived, we relaxed the assumption that the $RV$ for both components is equal to $-$3~km\,s$^{-1}$ ($RV$ of HD~144432~A, Table~\ref{tbl:stellparam}) and shifted their spectra until the Gaussian fit to the cross-correlation function has its center at 0~km\,s$^{-1}$. In this way, we obtained heliocentric $RV$s for components B and C of $-5\pm2$ and $-4\pm2$~km\,s$^{-1}$, respectively. In Fig.~\ref{fig:HD144432BC}, we display the observed FEROS spectrum, the final combined synthetic spectrum of components B, C, and A, and the relative flux contribution of each component. 

Finally, we note that to assume that the flux ratio of components B to C is equal to the $M_V$ ratio, is equivalent to assuming that they are at the same distance and have the same extinction. If we allow for a different brightness for both components, we find that a slightly closer fit to the 7700~\AA~ line can be achieved if C is slightly brighter (by up to a 0.5~mag). However, it is statistically very unlikely that two stars of similar $RV$, which indicates a common membership, and co-eval evolutionary status are so close together on the sky, thus it is unlikely that C is brighter than B.

From the measured distances between A, B, and C (Table~\ref{tbl:astro}) and their derived masses (Table~\ref{tbl:stellparam}), the components B and C orbit component A with a period of 2100$\pm$600~yr and B and C orbit each other with a period of 60$\pm$20~yr under the assumption of circular orbits and that the triple system is seen pole-on.
\subsection{Signatures of youth in HD 144432 B and C}
The lithium line at 6708~\AA ~is visible in the combined spectrum of HD 144432 B and C. Because the lithium line of the B+C observed spectrum is stronger (EW 0.52$\pm$0.02\AA) than the lithium line expected for the combination of a K7V and a M1V spectra (EW 0.42$\pm$0.02\AA), we conclude that HD~144432 B and C are likely young stars.

To constrain the age of the companions from the position in the HRD (Fig.~\ref{fig:hrd}), we estimated their luminosities. We used MARCS stellar models \citep{gus08}, $K_{s}$-band photometry of the NACO images (Sec.~\ref{sec:imaging}), and assumed a constant extinction of A$_{V}$=0.15 (Sec.~\ref{sec:stellparamsA}) for all components. We derived a luminosity of $L_{\rm B}=0.50^{+0.24}_{-0.20}$~L$_{\odot}$ for the B component and $L_{\rm C}=0.41^{+0.19}_{-0.16}$~L$_{\odot}$ for the C component. The derived ages and other derived stellar parameters are listed in Table~\ref{tbl:stellparam}. The ages of the three stars are comparable taking into account the error bars. Therefore, we estimate that the age of the system is 6$\pm$3~Myr (weighted mean of the individual age values), which is in good agreement with the value of $8^{+3}_{-1}$~Myr derived by \citet{car07}.
\section{Discussion and conclusions}\label{sec:discussion}
We have analyzed NIR AO images obtained with NACO of the young Herbig star HD~144432 and discovered that it is a triple system. The components B and C are located 1.47\arcsec~north of the primary star, and B and C themselves are separated by 0.1\arcsec. Using high-resolution optical spectra, we have determined the stellar parameters $T_{\rm eff}$, $\log g$, $L_{\star}$, $M_{\star}$, and $v\sin i$ of all three components and derived a spectral type of A9/F0Ve for component A and K7V and M1V for components B and C, respectively. This is in agreement with the trend suggested by \citet{cor98} that companions of Herbig Ae stars are usually of spectral type K to M.

The derived ages from the position in the HRD of all components are comparable within the error bars and the system age of the triple system can be constrained to be 6$\pm$3~Myr. This implies that these stars have a common evolutionary state and therefore supports the scenario of formation via cloud fragmentation.
\begin{acknowledgements}
We are very grateful to the referee for comments and suggestions that helped to improve the paper. This research has made use of NASA's Astrophysics Data System Bibliographic Services.
\end{acknowledgements}
\bibliographystyle{aa}
\bibliography{refs}
\listofobjects
\end{document}